\documentclass[iop]{emulateapj}
\usepackage{graphicx}
\usepackage{amsfonts}

\newcommand{\kms}{km~s$^{-1}$} 
\newcommand{\etal}{et al.}
\newcommand{\gal}{$\alpha$}

\begin{document}
\shorttitle{NGC 1866}
\shortauthors{Dupree et al. }

\title{NGC 1866: First Spectroscopic Detection of Fast Rotating Stars\\
 in a Young LMC Cluster}

\author{A. K. Dupree }
\affiliation{Harvard-Smithsonian Center for Astrophysics, 60 Garden Street, Cambridge, MA 02138, USA}
\author{A. Dotter }
\affiliation{Harvard-Smithsonian Center for Astrophysics, 60 Garden Street, Cambridge, MA 02138, USA}
\author{C. I. Johnson }
\affiliation{Harvard-Smithsonian Center for Astrophysics, 60 Garden Street, Cambridge, MA 02138, USA}
\author{A. F. Marino}
\affiliation {Australian National University, The Research School of
Astronomy and Astrophysics, Mount Stromlo Observatory, AU}
\author{A. P. Milone}
\affiliation {Australian National University, The Research School of
Astronomy and Astrophysics, Mount Stromlo Observatory, AU}
\author{J. I. Bailey III} 
\affiliation{Leiden Observatory, Niels Bohrweg2, 2333CA Leiden, NL}
\author{J. D. Crane}
\affiliation {The Observatories of the Carnegie Institution for Science, 813 Santa
Barbara Street, Pasadena, CA 91101, USA}
\author{M. Mateo}
\affiliation {Department of Astronomy, University of Michigan, Ann Arbor, 
MI 48109, USA}
\author{E. W. Olszewski}
\affiliation {The University of Arizona, 933 N. Cherry Avenue, Tucson, AZ 85721, USA} 

\accepted{10 August 2017}

\begin{abstract}
High-resolution spectroscopic observations were taken of 29 extended main sequence  
turn-off (eMSTO)   stars in the young ($\sim$200 Myr) LMC cluster, NGC 1866  
using the Michigan/{\it Magellan} Fiber System and MSpec spectrograph on the {\it Magellan}-Clay 
6.5-m telescope. These spectra reveal the first direct detection
of rapidly rotating stars whose presence has only been inferred from photometric studies. 
The eMSTO  stars exhibit H\gal\ emission 
(indicative of Be-star decretion disks), others have
shallow broad H\gal\ absorption (consistent with rotation
 $\ga$150 \kms),   or deep H\gal\ core absorption
signaling lower rotation velocities ($\la$150 \kms). The spectra appear consistent
with two populations of stars - one rapidly rotating, and the other,  younger and  slowly rotating.

\end{abstract}

\section{Introduction}
Identification of multiple main sequences in old Milky Way  globular clusters
from  HST photometry (Bedin et al. 2004; 
Piotto \etal\ 2007; Gratton \etal\ 2012) created 
a  fundamental change in our concept of their stellar populations 
for it suggested that cluster stars are neither coeval nor chemically
homogeneous. This  paradigm shift  results from the fact that
multiple sequences are visible along the entire color-magnitude diagram (CMD) 
signaling two or more generations of stars.

No completely successful  scenario exists to explain multiple populations although  
many possibilities have been offered.  The most popular suggests that 
a second generation of enriched (polluted) stars forms from gas that
was processed at high temperatures in the cores and/or envelopes of intermediate
to high mass first generation stars.  Each of the many possibilities
appears to have at least one fatal flaw (Bastian \etal\ 2015; Renzini \etal\ 2015; 
Charbonnel 2016). The situation becomes more complicated when investigating younger clusters, which 
could reveal the predecessors of the Milky Way clusters. 

Photometric studies of young and intermediate age clusters (age $<$2 Gyr) in the 
Large Magellanic Cloud (LMC)  support yet another scenario. They have revealed 
an extended (broadened)  main-sequence turnoff (eMSTO)  and/or a bimodal main sequence
(Mackey \etal\ 2008; Milone et al. 2009; Goudfrooij \etal\ 2009, 2014).  This
discovery could imply that a prolonged (100-500 Myr) 
star-formation history occurred  (Mackey \etal\ 2008; Conroy \& Spergel 2011; Keller \etal\ 2011). This could
be an attractive simple explanation since there are 
concerns about the lack of active star-formation
in clusters older than 10 Myr (Niederhofer \etal\ 2016) and the
absence of natal cluster gas after 4 Myr (Hollyhead \etal\ 2015) suggesting
that multiple stellar generations may not be present.  

Photometry of young ($\sim$300 Myr)
stellar clusters also reveals the eMSTO and a bifurcated main sequence  
(D'Antona \etal\ 2015; Milone \etal\  2016, 2017). A recent claim 
of detection of young stellar objects in some  young clusters in the LMC  
hints at ongoing star formation (For \& Bekki 2017).  However, other scenarios
have been introduced to explain the eMSTO and bifurcated main sequence
including a range of ages (Mackey \& Broby Nielsen 2007; Milone \etal\ 2009; Keller \etal\ 2011; 
Correnti \etal\ 2014; Goudfrooij \etal\ 2014), different rotation rates (Bastian \& deMink 2009, 
Bastien \etal\ 2016; Niederhofer \etal\ 2015; D'Antona \etal\ 2015; Milone \etal\ 2016, 2017),
braking of rapid rotators (D'Antona \etal\ 2017), or
different metallicities (Milone \etal\ 2015).

Our target, NGC 1866, a 200 Myr cluster in the LMC, displays the eMSTO and also a bifurcated main
sequence (Milone \etal\  2017).  
These characteristics are not due to photometric errors,
field-star contamination, differential reddening, or non-interacting binaries (Milone \etal\ 2016, 2017).
Comparison with isochrones (Milone \etal\ 2017) suggests that the best-fit
of the bifurcated main sequence  comes from rotating stellar 
models for the red main sequence and non-rotating models for the blue main sequence. 
It is believed that abundances are similar among the populations of NGC 1866 
(Mucciarelli \etal\ 2011), although
the ages are not, and may range from 140 Myr to 220 Myr (Milone \etal\ 2017).  Isochrone 
modeling
provides good agreement with the main-sequence objects but the fit to the eMSTO
objects is not as satisfactory.   Variable stars,
such as $\delta$ Scuti objects might also produce an extended turn off (Salinas \etal\ 2016),
however these stars become significant in older clusters (1$-$3 Gyr) where the 
turnoff from the main sequence coincides with the instability strip.  
Stellar rotation not only affects the colors of the stars but also
their lifetimes through rotational mixing. Possibly rotation could cause the
observed spreads in the CMD (Bastian \& deMink 2009). In fact,  
narrow and broad-band photometry of bright stars in two young  LMC clusters 
hints at the appearance of H\gal\ emission
(Bastian \etal\ 2017) which is interpreted as signaling the presence of rapidly rotating
Be stars.

No direct  measure of rotation  has
been carried out for individual stars populating the eMSTO in  LMC clusters.  
In this paper, we report the first high-resolution
spectroscopy  of the H\gal\ line in  29 stars in the extended turnoff region of the  LMC
cluster NGC 1866.  Synthesis of model spectra indicated that narrow
photospheric features would be `washed out' and too subtle to detect if the stars
are rapidly rotating, making the H\gal\ transition a feature of choice to characterize
the rotational state of the star.

\section{Spectroscopic Material} 

Stellar spectra  were obtained with the Michigan/{\it Magellan} 
Fiber System (M2FS, Mateo \etal\ 2012) 
and the MSpec multi-object spectrograph  mounted on the {\it Magellan}-Clay
6.5-m telescope at Las Campanas Observatory.  The fibers have
a diameter of 1.2'' and can span a field of view nearly 
30 arcminutes in diameter. A  180$\mu$m slit yielded a resolving 
power $\lambda /\Delta \lambda \sim 28,000$. The spectra were binned
by 2 pixels in the spatial direction and remained at 1 pixel along
the dispersion. 
  
The selected targets, which are likely  members of the cluster NGC 1866,  
were identified by Milone et al. (2017)
from the Ultraviolet and Visual Channel of the Wide Field Camera 3 
(UVIS/WFC3) of HST.  Images  taken with the F336W filter 
and the F814W filter provided the photometry and astrometry.  
Milone \etal\ (2017) 
noted that the apparent stellar density became constant at radial distances
greater than about 3 arcminutes from the cluster center, and concluded that
cluster members did not extend beyond that distance.  Our targets comply
with this criterion.   In addition, we selected targets 
separated by 2.5 arcsec at a minimum from any neighboring stars 
that are brighter and located away from stars less than  2 magnitudes 
fainter in the F814W band than the target star.  With this
selection criterion, coupled with the requirements on fiber placement, 
very few stars remain within the half-light radius
of the cluster (41 arcsec); in fact only two of our targets are located there.
The vast majority lie between 41 arcsec and $\sim$180 arcsec from the center.
This criterion identified  $\sim$ 150 acceptable targets, spanning
V = 16.2--20. Positions of the guide and acquisition stars  were 
verified by comparison with the 2MASS catalog and WFI images. 
The software code for M2FS fiber positioning 
selected targets according to our priorities. 

We chose the filter ``Bulge-GC1'' which
spans 6120--6720\AA\ over 6 echelle orders, and allows up to 48 fibers
to be placed on our targets.  In practice, several fibers are
placed on the sky; thus we obtained about 43 stellar targets 
per configuration. Some targets were ``lost'' due to low fiber sensitivity,
neighboring very bright stars, or possibly inaccurate coordinates.  
Two configurations - a bright and faint  selection --
each spanning about 2 magnitudes were implemented. Our principal configuration
was observed on 8 December and 12 December 2016 with  7 exposures totaling 5.5 hours
varying between 2100s and 2700s each.  A fainter target configuration was observed
on 11 December and 13 December 2016, but the spectra were severely compromised 
by the full moon. 

Standard IRAF procedures  performed the bias subtraction,
overscan trimming, and combination of the 4 individual CCD quadrants into
one monolithic array. The {\it dohydra} task was implemented for 
aperture identification and tracing, scattered-light subtraction,
flat-fielding, and line identification for  wavelength calibration 
from the ThAr comparison lamp. Sky emission lines were identified
and removed individually and the H$\alpha$ order was continuum
normalized with a cubic spline, omitting the H\gal\ region. A detailed
description of the procedures can be found in 
Johnson \etal\ (2015).\footnote{An outline of  procedures is available online ({\it
https://www.cfa.harvard.edu/oir/m2fsreduction.pdf}).}    We obtained 
H\gal\ spectra for 29 targets within
a 3 arcmin radius of the cluster center. Comparison of stars
in the reference field within the color and magnitude boundaries of
our sample suggests that $~$10\% of our targets (comprising $~$3 targets) in
the cluster field might not be cluster members.   The solar H\gal\ line in absorption  
frequently appears in the spectra at shorter wavelengths than the LMC spectral 
features but nevertheless allows definition of the continuum on the
short wavelength side of H$\alpha$.  Target stars, their positions, 
magnitudes,  and H\gal\ characteristics are given
in Table 1.

\section{Analysis}
H\gal\ spectra of the 29 targets located within 3 arcmin of the 
cluster center are shown in Fig. 1 where both emission
and  absorption can be found.  The emission features  
are centered on the velocity of the cluster, 
$+$298.5 \kms\ (Mucciarelli \etal\ 2011).  The profiles are typical
of those found in Be stars in which the
emission arises in a Keplerian decretion disk surrounding
a rapidly rotating star (Rivinius \etal\ 2013; Paul \etal\ 2017; 
Reid \& Parker 2012).  Differences 
in the profile shapes result from the angle of observation, from pole-on to equator (Struve 1931). 
In particular, the narrow `wine-bottle' H\gal\ profile of Object 58 suggests it is viewed nearly
pole-on; many others (Object 14, 26, 56, 62, 89) exhibit a deep central
absorption thought to arise from absorption in the cool circumstellar disk when viewed
edge-on (Hummel 1994).
The line widths at the continuum level vary as well, from $\pm$110 \kms\ in the
pole-on object to $> \pm200$ \kms\ in stars observed at intermediate angles.

Absorption profiles shown in Figure 1  are shallow and broad for many stars. 
H\gal\ absorption wings in B stars are indicative of the stellar gravity, and
the core of the line  responds to rotation, becoming more shallow
with increasing values of $v \sin i$. Several of the stars 
can be seen by visual inspection to have a deep (narrow) core in the absorption profile.

We further examine the absorption profiles in two ways: profile synthesis and 
broadening assessment. 
In the first instance, theoretical H\gal\ absorption profiles are
compared to the observed  profiles for  three bright targets in Fig. 2a, 2b, 2c.  
H\gal\  profiles from  Castelli and Kurucz LTE 
models\footnote{Available at {\it http://kurucz.harvard.edu.}}  were broadened 
using a Gaussian convolution and overlaid on the profiles; also shown are reasonable excursions
to the profile with higher and lower rotational velocities.  Comparison of 
LTE vs non-LTE calculations
of H\gal\ profiles shows that LTE profiles are adequate for stars cooler than $\sim$22,000K 
(Przybilla \& Butler 2004; Nieva \& Przybilla 2007).  HST colors predicted from 
the Choi et al. (2016) models suggest these targets have $T_{eff}\sim$15,000K. Synthesis of the  
spectra for Object 10 and 12 suggests $v \sin i \sim $ 70$-$100 \kms\ in contrast to a value 
$\sim$200 \kms\ for Object 30.  These values
provide a lower limit to the true velocity because the orientations of the stars are unknown.  
Secondly,  we developed a broadening parameter defined as the
ratio of the central depth to the line profile depth at a wavelength 4\AA\ longward of line center:
$R_{c+4}$. This ratio was measured for the target stars after subtracting the solar scattered continuum. Model
profiles demonstrate that this ratio increases with increasing velocity (Fig. 2d).  The
dependence of the ratio on velocity appears similar for different values of the gravity.  
The observed profiles map velocities, {\it v}, from 50 to 250 \kms.
Inspection of the H\gal\ absorption
profiles suggests that the majority of the targets with {\it v} $\la$ 150 \kms\ exhibit deep
absorption cores, therefore we denote these stars as `slow rotators' and label the
stars with {\it v} $\ga$150 \kms\ as `fast rotators'.

Theoretical critical velocities are shown in the HST CMD  
from MIST isochrones (Dotter 2016; Choi \etal\ 2016)
for a range of ages corresponding to NGC 1866 (Fig. 3). The isochrones have been
shifted by the assumed distance and reddening of NGC 1866 [($m-M)_0$= 18.31, $E(B-V)$= 0.11, 
Milone \etal\ 2017]. The MIST isochrones include the effects of rotation; those 
shown in Fig. 3 are initialized
with $\Omega/\Omega_{crit}=0.4$ at the ZAMS.   
The critical velocities for these targets (350$-$400 \kms) are
larger than the values inferred from Fig. 2d.
This may account for the lack of emission in H\gal\ as stars have not achieved
velocities necessary to shed material producing emission from a surrounding disk.

\section{Discussion}
The majority of eMSTO target stars  fall into two categories: fast and slow rotators.
Detection of H\gal\ emission clearly signals a fast-rotating star with
a Keplerian decretion disk - the Be phenomenon (Rivinius \etal\ 2013). We
do not yet have measurements of the rotational velocity of the emission objects. 
The H\gal\ absorption profiles  indicate both rapidly and slowly 
rotating stars.  Isochrone fitting to HST cluster photometry (Milone \etal\ 2017) suggested 
that the blue stars on the bifurcated main sequence are 
slowly rotating, and represent  two  stellar
generations of 140 Myr and 220 Myr.  Red main sequence stars are believed to
be rapidly rotating ($\Omega$=0.9$\Omega_{crit}$, a high fraction of the
critical rotation rate, $\Omega_{crit}$), and are consistent with an age of 200 Myr.
Thus our spectroscopic results confirm the conclusion of Milone \etal\ (2017) from
photometry that identified  
fast and slowly rotating populations.

Figure 4 shows the HST CMD of NGC 1866 marked with 
targets and their characteristics. 
Inspection  suggests that two  targets, Object 3 and 36, 
are outliers, and perhaps not cluster members because their 
colors are $\lesssim -$1
and cluster isochrones (Fig. 3) do not extend to those colors.
We exclude them from the calculation of median parameters. Taking
the ``narrow" absorption targets as those with $R_{c+4} \le 0.90$, 
corresponding to {\it v} $\la$150 \kms, in 
comparison to the targets with emission, we find that  
the median magnitudes, m$_{F814W}$ are essentially identical:
18.01$\pm$0.418 (narrow) and 17.99$\pm$0.377 (emission). Here the 
dispersion is calculated as the median of the absolute deviations
of magnitude about the median magnitude.  However, the median colors
suggest what is evident from Figure 4, namely $m_{F336W}-m_{F814W}$
equals $-$0.56$\pm$0.10 (narrow) and $-$0.34$\pm$0.12 (emission).
Targets exhibiting broad absorption $R_{c+4} > 0.90$ have
a median magnitude similar to the others, and a median color
lying between the values of the other groups: $-$0.50$\pm$0.10. 

Additional characteristics of the stars  can be
compared to the results of the photometric studies 
of the cluster.  Three results derive from the photometry:

\smallskip

(1) {\it The red main sequence (rapid rotators) is more centrally concentrated than the blue
main sequence (slow rotators).}  Milone \etal\ (2017) find the fraction of red 
main sequence stars within 1 arcmin of the cluster core 
is $\sim$0.68. Excluding outliers, the   5  targets within 1 arcmin include 4 stars which 
are  fast rotators, exhibiting 
H\gal\ emission  (corresponding to a fraction of  
0.80$\pm$0.53 using Poisson statistics).  Thus there appears to be  a preponderance 
of rapidly rotating stars in 
eMSTO objects located in the  core of the cluster.  Between 1 and 3 arcmin from the cluster 
center, our sample of 22 eMSTO stars indicates
the fast rotators decrease slightly to 0.55$\pm$0.20 of the targets at this distance -
although within the error estimate, the fraction remains comparable to that in the core.

\smallskip

(2) {\it The `blue component' comprises about 0.15 of the stars at the top
of the main sequence.}  We find the fraction of slow rotators (the blue
component) to be 0.41 in the total eMSTO sample  -- a value higher than the photometric results.

\smallskip

(3) {\it Isochrone fitting suggests that the fast rotating population has
a velocity of 0.9 of critical velocity and corresponds to $\sim$ 200 Myr; the non-rotating
isochrones indicate that the blue main sequence may harbor two populations
of 140 Myr and 220 Myr.}  We find the rapid rotators  occur to the
red of the slowly rotating stars in the HST CMD (Fig. 4). 
Inspection of isochrones
(Milone \etal\ 2017) shown in  Fig. 4, suggests that the slowly
rotating objects span the non-rotating isochrones between 140 Myr and slightly less than $\sim$220 Myr.
The rotating population lies to the `red' of  the slower-rotating stars which 
suggests a population older than 200 Myr if $\Omega \sim 0.9\Omega_{crit}$ and 
perhaps comparable to the
slowly rotating objects.  If so, this would remove the uncomfortable problem presented 
from photometry (Milone \etal\ 2017) of three populations harboring slow, fast, and slowly
rotating stars formed in sequence.

\vspace{0.3 in}
  
Stars exhibiting H\gal\ emission comprise a fraction 0.41 of our total sample.  This value 
is comparable to the fraction (Bastian \etal\ 2017)
inferred from narrow and broad-band photometry of the eMSTO  spanning 0.4-0.62 in the young LMC 
clusters NGC 1850 (79 Myr) and 0.33 in NGC 1856 (300 Myr). Photometric studies,
however, give a lower limit to the emission fraction since only stars with strong
emission are detected.  Moreover, radial velocity shifts of LMC clusters
can be significant and compromise the detection of H\gal\  in the narrow HST filter
F656N frequently used as an H\gal\ diagnostic. Spectroscopy is advantageous
for H\gal\ detection because  weak emitters can be identified, 
rapid rotators without H\gal\ emission can be detected, and radial velocity shifts
are of no consequence.   Inclusion of stars with broad H\gal\ absorption 
raises the rapid rotation fraction to 0.61 among the eMSTO population in 
our total sample and implies the fraction must be higher in other clusters, e.g. NGC 1850 
and NGC  1856, as well.

Direct spectroscopic measures of the eMSTO stars clearly demonstrates the
presence of rapidly rotating stars that are cooler than  a  population of slowly rotating
objects. It is not understood how such conditions were established. 
If the populations were coeval, slowly rotating stars evolve faster than the
rapid rotators and they  should have a lower turnoff luminosity. The CMD 
of the eMSTO objects (Fig. 4) displays the opposite behavior which argues for 
an actual  spread in age: 
the rapidly rotating population marks the (older)  initial burst of star 
formation, followed by a second generation that is more slowly rotating. Isochrones
in Fig. 4 demonstrate the younger non-rotating objects (at 140 Myr) lie to
the 'blue' of an older (200 Myr) isochrone that has a rotation close
to the critical velocity, here $\Omega = 0.9 \Omega_{crit}$.   
Recently, D'Antona et al. (2017) have speculated that rotational braking might mimic 
an age spread, a conjecture which requires spectroscopic confirmation 
by abundance measures or detection of a stellar wind.
The spatial distribution is also puzzling.   
Goudfrooij \etal\ (2011) find  that the upper eMSTO (presumably younger objects)  
is significantly more centrally 
concentrated than the lower eMSTO in many massive intermediate age clusters in the LMC. 
This is in harmony with  a second generation of stars formed from material shed
by stars of the first generation.  Our results might suggest a different scenario.
It is the cooler eMSTO objects, spectroscopically determined
to be fast rotators, that dominate within 1 arcmin of the cluster core, although
we have a small sample.  Perhaps this is 
typical of less massive and/or younger clusters. 
Yet, it is puzzling that the rapidly rotating stars are concentrated towards the
cluster center (Milone et al 2017) where it might be expected 
that a second stellar generation
would form from the material of the first generation. This would appear to suggest
that another scenario must be sought for young clusters.

\section{Acknowledgments} We thank the anonymous referee for  thoughtful comments and
advice on the manuscript. AFM and APM acknowledge support by the
Australian Research Council through Discovery Early Career Researcher Awards 
DE160100851 and DE150101816. EWO was partially supported by NSF Grant AST1313006.  
This research has made use of NASA's 
Astrophysics Data System Bibliographic Services. 
And we have used  data products from 2MASS, 
which is a joint project of the University of Massachusetts 
and  IPAC/Caltech,  funded by  NASA and the NSF.

\medskip

\facility{Facility: Magellan: Clay (M2FS)}

\begin{figure}[ht!]
\hspace*{-0.3in}

\includegraphics[scale=0.7]{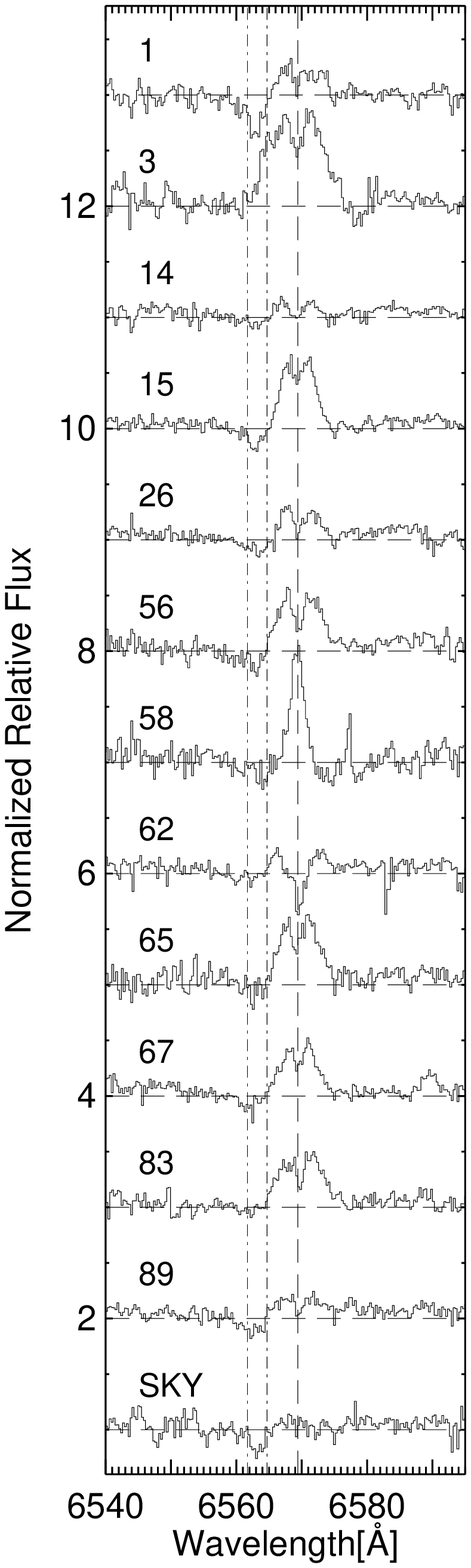}
\vspace*{-7.in}

\hspace*{+1.95 in}
\includegraphics[scale=0.7]{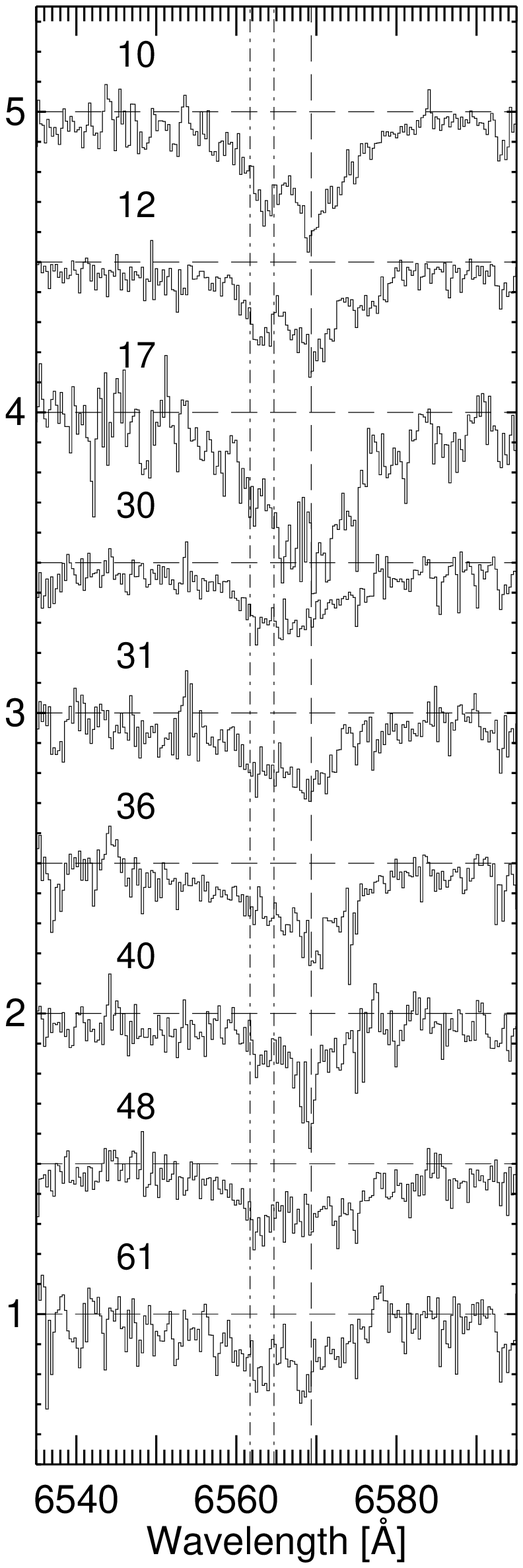}
\vspace*{-7.13in}

\hspace*{+4.3in}
\includegraphics[scale=0.7]{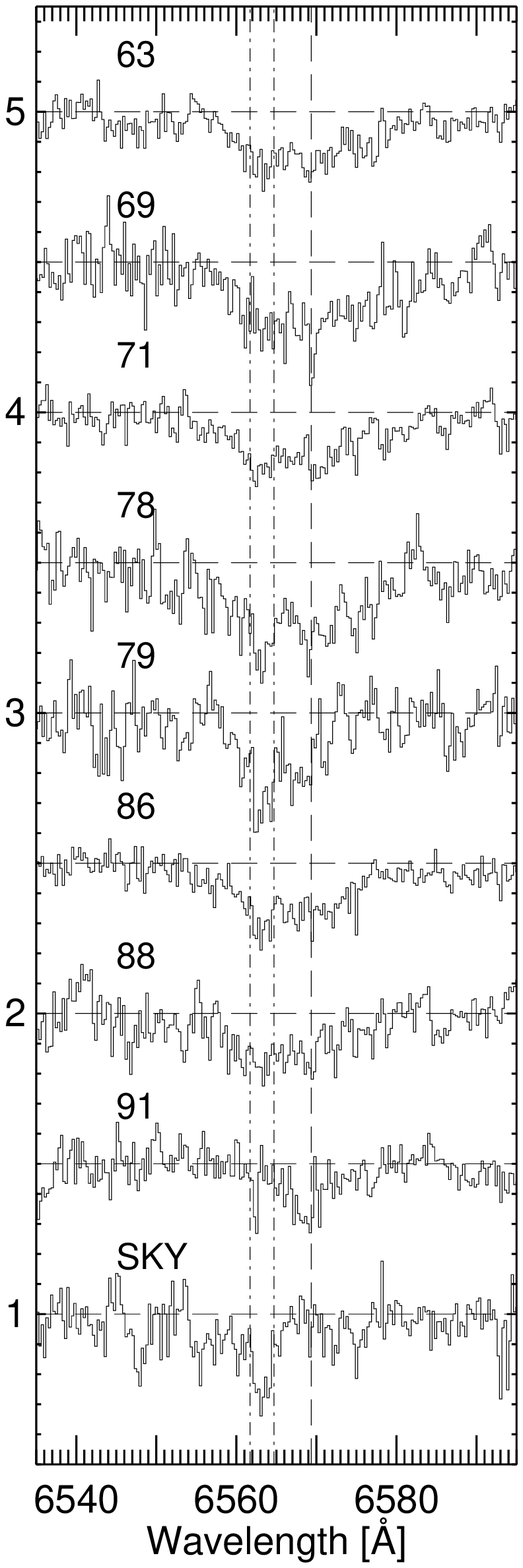}

\parbox{6.5in}{\caption{ The H\gal\ region of the  eMSTO stars in NGC 1866. 
Continuum-normalized 
spectra are binned to a  resolution element and offset. 
The heliocentric velocity of the cluster, $+$298.5$\pm$0.4 \kms\ 
(Mucciarelli \etal\ 2011) is indicated by
the broken  line.  The dot-dash lines mark the position
of the scattered solar H\gal\ line.
A deep sky spectrum is shown. 
Object numbers are marked. {\it Left panel:} H$\alpha$ 
emission objects. {\it Center and Right panels:} Stars exhibiting H\gal\ absorption.  }}

\end{figure}


\begin{figure}                                             

\includegraphics[scale=0.35, angle=90]{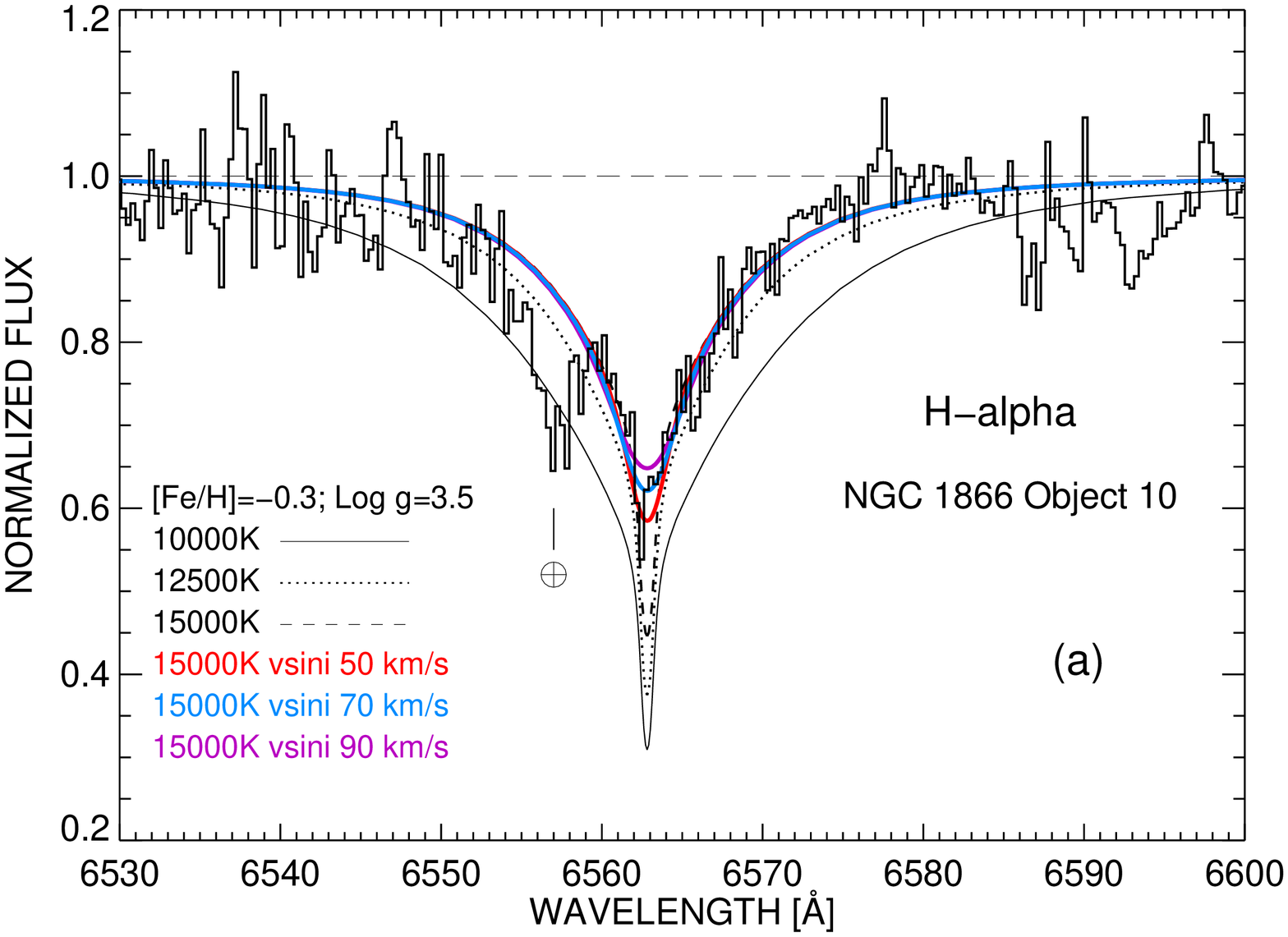}
\includegraphics[scale=0.35, angle=90]{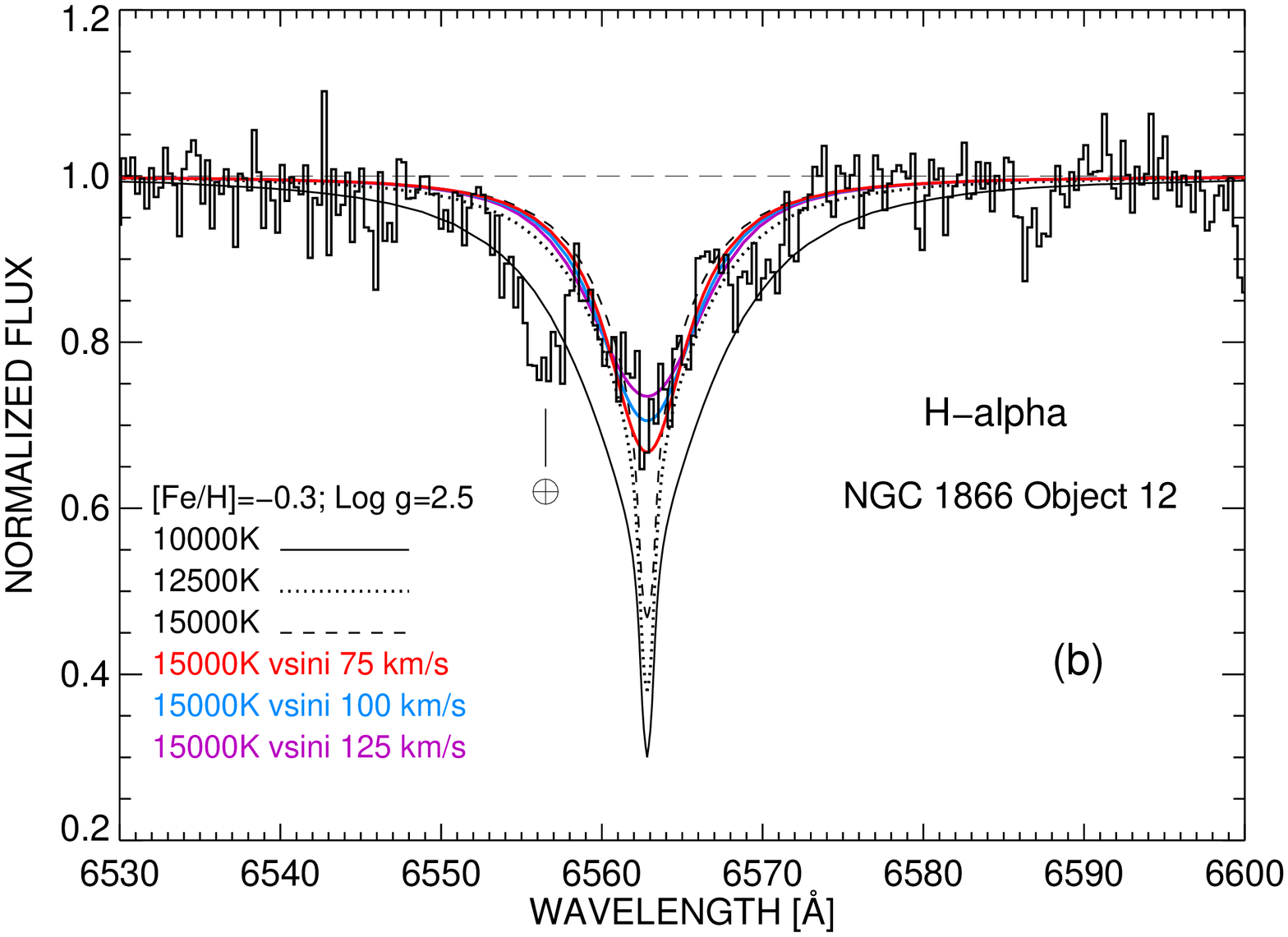}

\includegraphics[scale=0.35, angle=90]{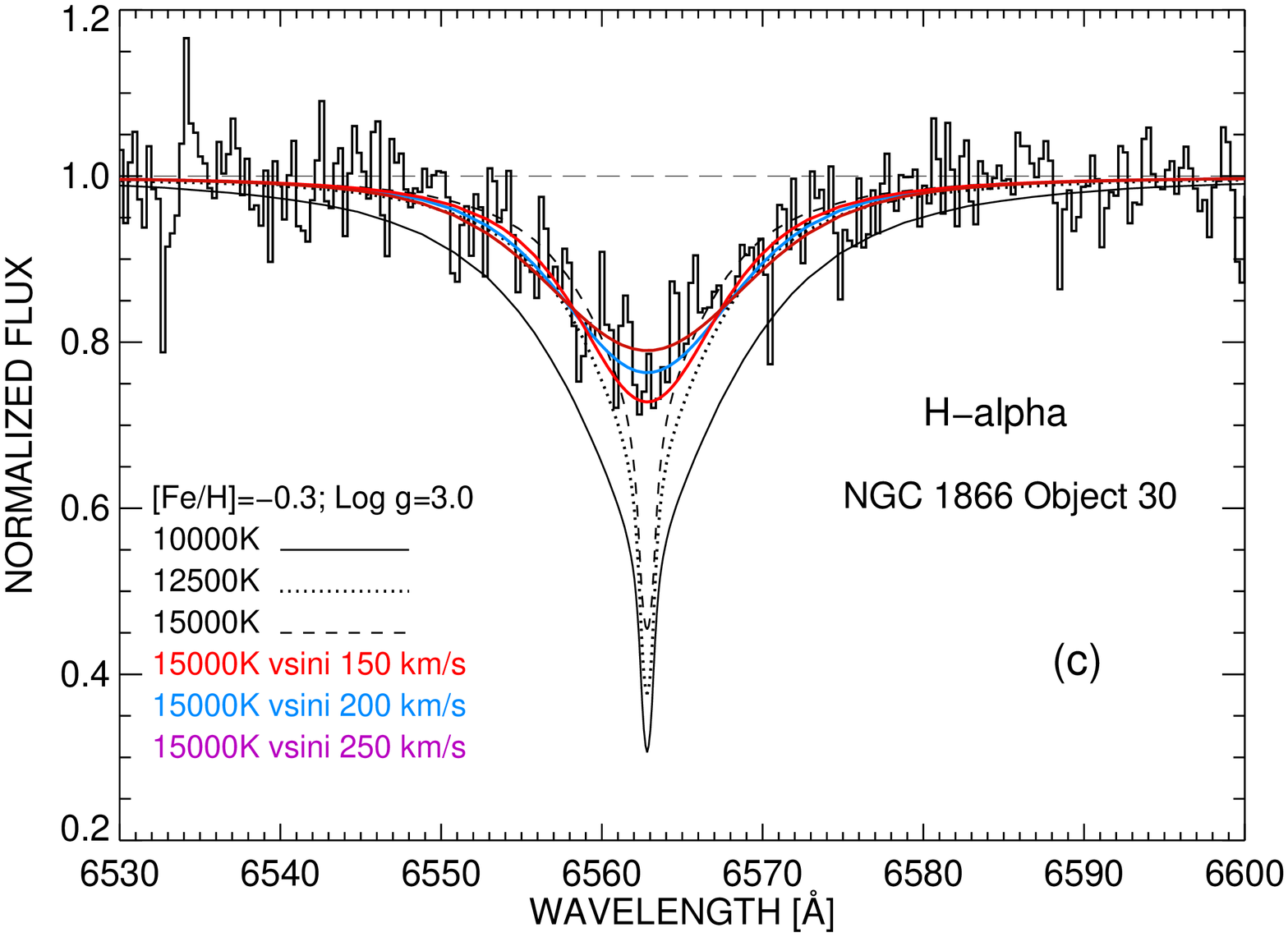}
\includegraphics[scale=0.35, angle=90]{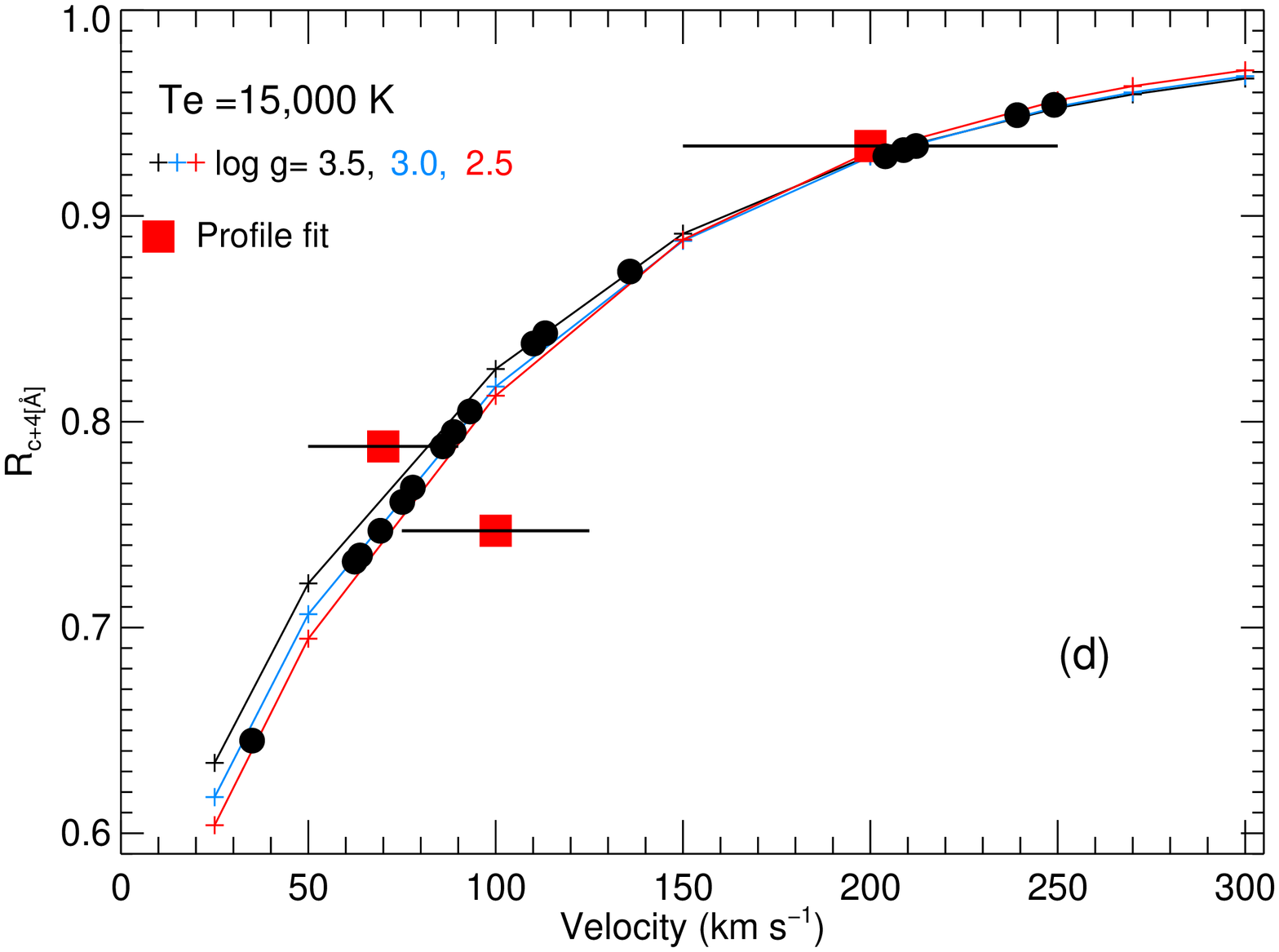}

\caption{{\it Panels a, b, c:} The H\gal\ region in  three objects exhibiting
absorption. The spectra are centered on H\gal. H\gal\ 
sky contamination is marked in Object 10 and 12.  
Theoretical profiles  are  from {\it kurucz.harvard.edu} for 
[Fe/H]=$-$0.3,  {\it log g} values of 2.5 $-$3.5, and  broadened by a Gaussian 
to simulate rotation. The profiles
suggest values of {\it v~sin~i} in excess of 50 \kms\ and ranging to $\sim$ 250 \kms.
Approximate upper and lower limits of velocities are shown.
Object 10 and Object 12, exhibiting deep narrow H\gal\ are considered to be slow rotators 
as compared to Object 30.
{\it Panel d:} Models of the relative central depth of H\gal\ ($R_{c+4}$) as compared to model fits
in panels {\it a, b,} and {\it c} are denoted by red squares.  The measured values of $R_{c+4}$ for the targets 
are marked by black circles.}

\end{figure}

\begin{figure}
\vspace*{2in}
\includegraphics [scale=0.7]{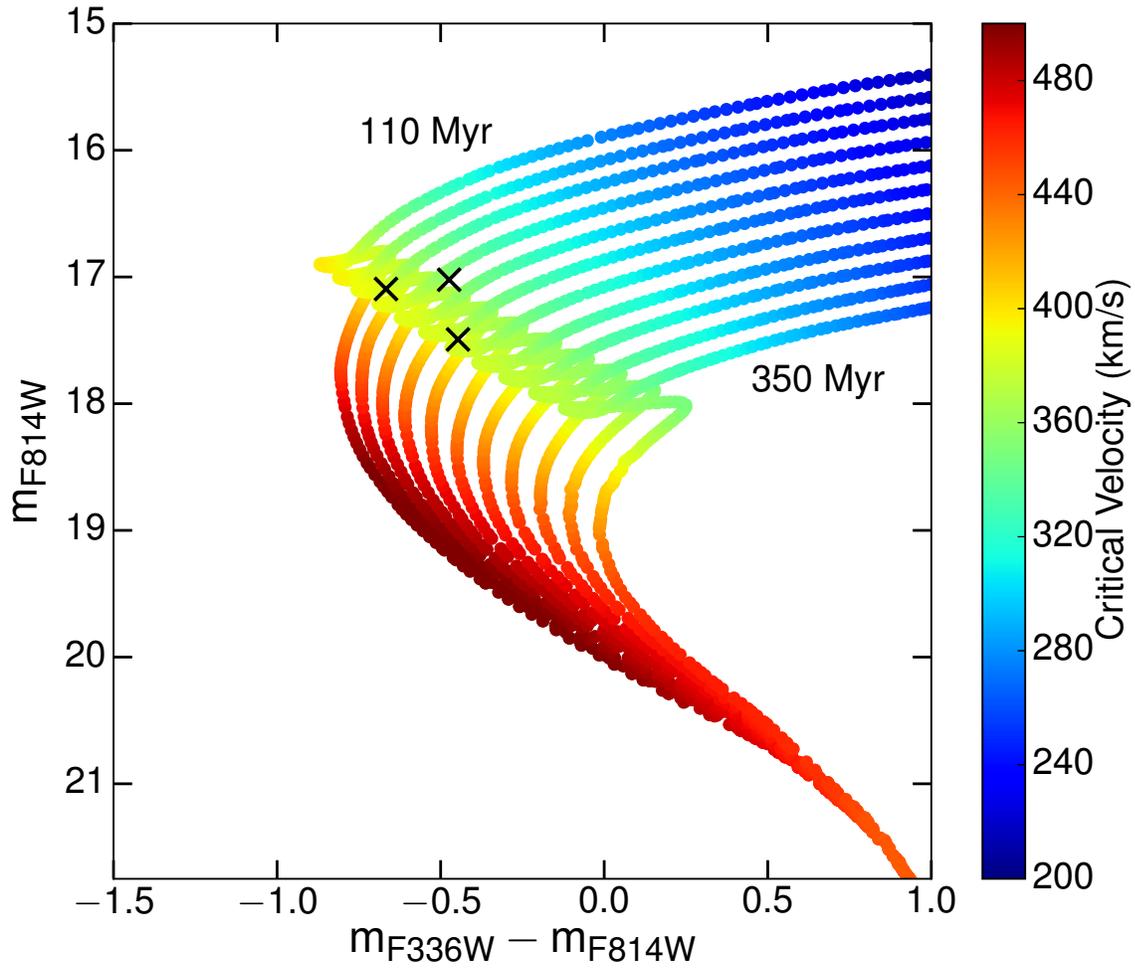}

\caption{Expected breakup rotational velocities for ages spanning 110  to 350 Myr. 
Positions of the three stars from Figure 2 (Object 10, 12, 30) are marked and
span isochrones from 140 to 225 Myr.  }
\end{figure}

\begin{figure}[ht!]
\begin{center}
\includegraphics[scale=0.6, angle=90]{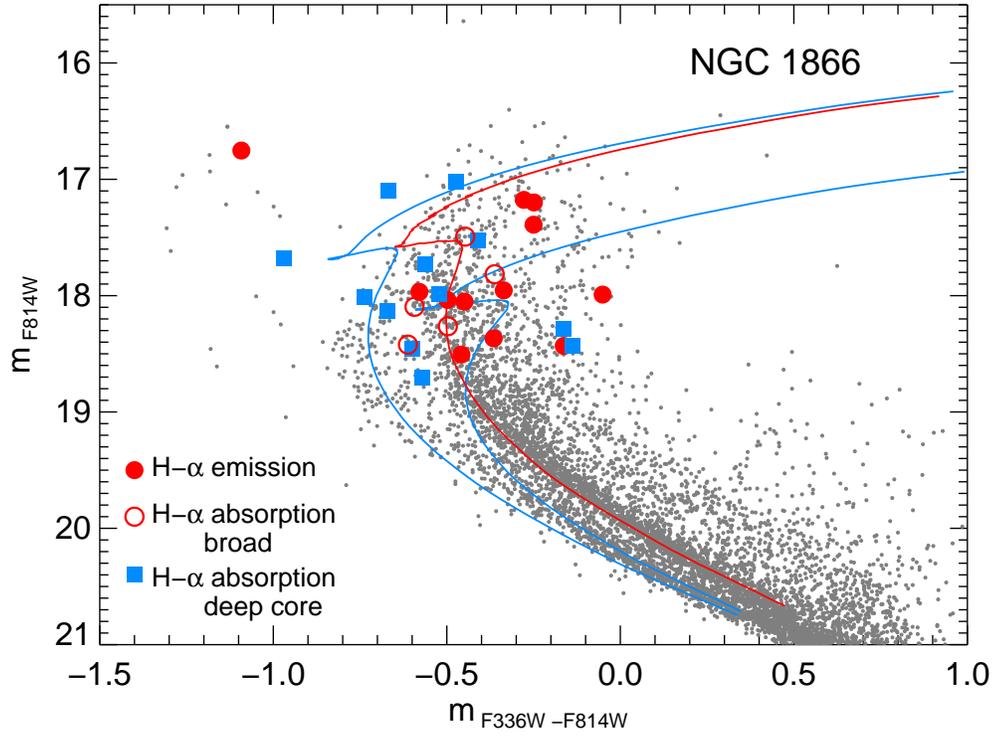}

\caption{Characteristics of  H\gal\ profiles  in NGC 1866 detected in the targets 
and displayed in the HST CMD. Isochrones are taken  from Georgy \etal\ (2013) for non-rotating
models ($\Omega = 0$) and  ages of 140 Myr and 220 Myr ({\it blue curves}) and
a rotating model ($\Omega = 0.9\Omega_{crit}$) with an age of 200 Myr ({\it red curve}) similar to those
shown by Milone \etal\ (2017). Stars 
within 3 arcminutes from cluster center are marked by grey dots. }
\end{center}
\end{figure}

\begin{deluxetable}{lcccrrlrccl}
\tablecolumns{11}
\tablewidth{0pt}
\tablenum{1}
\tablecaption{NGC 1866 Targets}
\tablehead{
\colhead{Object}&     
\colhead{RA (2000.0)}&
\colhead{DEC (2000.0)}  &     
\colhead{$m_{F336W}$}  &
\colhead{B     }   &
\colhead{V     }   &
\colhead{$m_{F814W}$}&
\colhead{Color\tablenotemark{a}}&
\colhead{d(arcmin)\tablenotemark{b}}&
\colhead{ R$_{c+4}$\tablenotemark{c} }&
\colhead{Key\tablenotemark{d}}}
\startdata
   1\tablenotemark{e} & 78.374466  & -65.481554 & 18.048 & 18.670 & 18.468 & 18.506& -0.458 & 1.39& ... & 1\\
   3 &   78.431151 &  -65.470138  & 15.661 &   16.732 &   16.843 &   16.753 & -1.092 &0.58& ... & 1\\ 
  10 &   78.418615 &  -65.475645  & 16.548 &   16.937 &   17.019 &   17.022 & -0.474 &0.69& 0.79 & 2\\  
  12 &   78.470013 &  -65.419725  & 16.428 &   16.881 &   17.038 &   17.095 & -0.667 &3.05& 0.75 & 2\\
  14 &   78.425251 &  -65.454748   & 16.887 &   22.488 &   15.825 &   17.177 & -0.290 &0.67& ... & 1\\ 
  15 &   78.369840 &  -65.465653  & 16.907 &   17.159 &   17.150 &   17.199 & -0.292 &1.06& ... & 1\\
  17\tablenotemark{f} & 78.477155 &  -65.468106  & 18.134 & 18.679 & 18.723 &  18.705& -0.571 &1.64& 0.87 & 2\\      
  26 &   78.393400 &  -65.455855  & 17.140 &   17.290 &   17.166 &   17.390 & -0.250 &0.69& ... & 1\\        
  30 &   78.360408 &  -65.473644  & 17.046 &   17.423 &   17.437 &   17.493 & -0.447 &1.40& 0.93 & 2\\     
  31 &   78.447226 &  -65.477066  & 17.114 &   17.551 &   17.493 &   17.525 & -0.411 &1.16& 0.73 & 2\\  
  36 &   78.449744 &  -65.441929  & 16.707 &   17.687 &   17.569 &   17.677 & -0.970 &1.65& 0.74 & 2\\    
  40 &   78.366140 &  -65.457688  & 17.164 &   17.633 &   17.639 &   17.727 & -0.563 &1.21& 0.64 & 2\\   
  48 &   78.459678 &  -65.436704   & 17.454 &   17.811 &   17.795 &   17.816 & -0.362 &2.04& 0.93 & 2\\
  56 &   78.409850 &  -65.475827  & 17.619 &   18.004 &   18.024 &   17.955 & -0.336 &0.69& ... & 1\\ 
  58 &   78.378488 &  -65.482908  & 17.389 &   17.864 &   17.910 &   17.968 & -0.579 &1.39& ... & 1\\
  61 &   78.354382 &  -65.457718   & 17.460 &   17.931 &   17.925 &   17.983 & -0.523 &1.49& 0.76 & 2\\
  62 &   78.424431 &  -65.442766   & 17.940 &   18.034 &   17.995 &   17.991 & -0.051 &1.34& ... & 1\\ 
  63 &   78.479936 &  -65.458286  & 17.272 &   17.911 &   17.987 &   18.010 & -0.738 &1.73& 0.77 & 2\\
  65 &   78.398687 &  -65.476448  & 17.539 &   17.927 &   17.932 &   18.037 & -0.498 &0.79& ... & 1\\     
  67 &   78.441385 &  -65.452302  & 17.603 &   18.019 &   18.084 &   18.052 & -0.449 &1.03& ... & 1\\
  69 &   78.495511 &  -65.440840  & 17.503 &   18.006 &   18.030 &   18.096 & -0.593 &2.52& 0.93 & 2\\
  71 &   78.463359 &  -65.451612  & 17.460 &   18.023 &   18.105 &   18.130 & -0.670 &1.49& 0.84 & 2\\
  78 &   78.457044 &  -65.466589  & 17.766 &   18.247 &   18.224 &   18.263 & -0.497 &1.13& 0.95 & 2\\ 
  79 &   78.473464 &  -65.449753  & 18.127 &   18.306 &   18.314 &   18.290 & -0.163 &1.77& 0.84 & 2\\  
  83 &   78.441541 &  -65.440489  & 18.002 &   18.392 &   18.355 &   18.367 & -0.365 &1.61& ... & 1\\ 
  86 &   78.510130 &  -65.437927   & 17.809 &   18.222 &   18.240 &   18.421 & -0.612 &2.92& 0.95 & 2\\
  88 &   78.395433 &  -65.444091  & 18.291 &   18.899 &   18.898 &   18.428 & -0.137 &1.29& 0.80 & 2\\
  89 &   78.452582 &  -65.456036   & 18.269 &   18.559 &   18.491 &   18.432 & -0.163 &1.13 & ... & 1\\
  91 &   78.443073 &  -65.430659  & 17.848 &   18.405 &   18.388 &   18.460 & -0.612 &2.17& 0.80 & 2          
\enddata
\tablenotetext{a}{Color defined by $m_{F336W}-m_{F814W}$.}
\tablenotetext{b}{Cluster center at RA(J2000): 05:13:38.92; DEC(J2000): -65:27:52.75 (McLaughlin 
  \& van der Marel  2005).}
\tablenotetext{c}{Ratio of depth of  H\gal\ line core to depth at  +4\AA\ in absorption line profiles.}
\tablenotetext{d}{1: H$\alpha$ 
emission; 2: H$\alpha$ absorption.}
\tablenotetext{e}{Spectrum from a second M2FS configuration:13 Dec. 2016 (UT),
total exposure  4.5 hours.}
\tablenotetext{f}{Spectrum from a second M2FS configuration:11 Dec. 2016 (UT),
total exposure  3 hours.}

\end{deluxetable}
\end{document}